\documentclass[preprint,12pt]{elsarticle}
\usepackage[utf8]{inputenc}
\usepackage{amsmath}
\usepackage{amsfonts}
\usepackage{amssymb}
\usepackage{graphicx}
\usepackage{cancel}

\let\today\relax
\makeatletter
\def\ps@pprintTitle{%
  \let\@oddhead\@empty
  \let\@evenhead\@empty
  \def\@oddfoot{\footnotesize\itshape
  {} \hfill\today}%
  \let\@evenfoot\@oddfoot
  }
\makeatother

\def\doubleunderline#1{\underline{\underline{#1}}}

\begin{document}

\begin{frontmatter}

\title{Impact of wall modeling on kinetic energy stability for the compressible Navier-Stokes equations}%


\author[label1]{Vikram Singh\corref{cor1}}
  \ead{vikram.singh@mpimet.mpg.de}

\author[label1]{Steven Frankel}
  \ead{frankel@technion.ac.il}

 \author[label2,label3]{Jan Nordström}
  \ead{jan.nordstrom@liu.se}

\cortext[cor1]{Corresponding Author, Present Address: Max Planck Institute for Meteorology, Hamburg, Germany}
\address[label1]{Faculty of Mechanical Engineering, Technion - Israel Institute of Technology, Haifa Israel}
\address[label2]{Department of Mathematics, Computational Mathematics, Linköping University,SE-581 83 Linköping, Sweden}
\address[label3]{Department of Mathematics and Applied Mathematics, University of Johannesburg, P.O. Box 524, Auckland Park 2006, South Africa}

\begin{abstract}

Affordable, high order simulations of turbulent flows on unstructured grids 
for very high Reynolds' number flows require wall models for efficiency. 
However, different wall models have different accuracy and stability properties. 
Here, we develop a kinetic energy stability estimate 
to investigate stability of wall model boundary conditions.
Using this norm, two wall models are studied,
a popular equilibrium stress wall model, which is found to be unstable  
and the dynamic slip wall model which is found to be stable. 
These results are extended to the discrete case using the Summation-by-parts (SBP) 
property of the discontinuous Galerkin method. 
Numerical tests 
show that while the equilibrium stress wall model is accurate but unstable,
the dynamic slip wall model is inaccurate but stable.
\end{abstract}



\end{frontmatter}

\section{Introduction}

High-order methods have been shown to provide more accurate solutions than low-order one's 
for the same number of degrees of freedom \cite{Hesthaven2008}.  
Here, we focus on the discontinuous Galerkin method (DG) 
\cite{osti_4491151, BASSI1997267, Cockburn2000}, 
which has become a popular tool for high order simulations 
for fluids over the past few decades. 
It can be made conservative, stable and arbitrarily 
high-order accurate on unstructured grids for linear problems.

There are studies indicating that DG is a suitable method 
for implicit large eddy simulations (ILES) \cite{Beck2014}. 
However, for wall resolved LES the total number of grid points required are of the order of
$\sim Re_L^{13/7}$, where $Re_L$ is the Reynolds' number based on streamwise length \cite{Choi2012}. 
The resources required to do simulations on such large grids are prohibitive. 
One way of avoiding this requirement is to use wall modeled LES (WMLES), 
where the no-slip boundary condition is replaced by a suitable model. 
In the case of WMLES, the grid points requirement scales as $\sim Re_L$ 
which is much more affordable.

Many different approaches exist for wall modeling \cite{LARSSON2016}. 
They can, in general, be categorized into algebraic and 
ordinary differential equation (ODE) based models. 
In the ODE based models, two grids are required.
One grid is used to solve the ODE governing the wall model
which provides the wall stress boundary conditions 
for the full set of equations. 
This procedure introduces an additional level of complexity. 
The algebraic models are simpler to implement requiring only solutions of 
algebraic equations.
Recent studies have shown the effectiveness of using such wall models for DG \cite{Frre2017}. 
Another more recent approach to wall modelling is the dynamic slip wall model 
\cite{Bose2014, Bae2018}.
It relies on deriving boundary conditions for the filtered Navier-Stokes equations.
This method requires a length scale as a 
regularization parameter which ensures that the no-slip
boundary condition is recovered as the resolution approaches 
direct numerical simulation (DNS) levels. 

Stability is highly desirable for any numerical method.  
Various methods exist to stabilize the discrete nonlinear equations
; see \cite{SVARD201417, Pirozzoli} and the references therein. 
However stability issues arising from the use of the wall model boundary conditions are not well studied.
If the boundary conditions of the initial boundary value problem (IBVP) are not well-posed, there is no guarantee that the simulation will be stable or even meaningful. 
Various studies have been done to determine boundary conditions that satisfy stability estimates \cite{SVARD20084805, BERG20117519, PARSANI201588}. 
However, these studies usually deal with standard boundary conditions 
such as no-slip, no-penetration or inflow/outflow type of boundary conditions.

In this paper, we consider two wall models. 
The first one is an equilibrium stress model and 
the second one a dynamic slip model.
Next, we derive kinetic energy stability estimates 
for the wall boundary conditions.
Using this bound, the equilibrium stress model is shown to be unstable 
while the dynamic slip wall model is stable.
We then introduce the DG method and the associated SBP property. 
The stability estimates for the wall models 
are then extended to the discrete case using SBP.
Finally numerical results are presented that compare 
the accuracy and stability of the two wall models.

\section{Wall models}\label{sec:wall}

There are various techniques for wall modelling in LES (WMLES) \cite{LARSSON2016}.
Here we will explore an equilibrium wall stress model 
\cite{Frre2017},
and the more recent dynamic slip wall model \cite{Bose2014}.

\subsection{Boundary conditions}

For the compressible Navier Stokes equations, four boundary conditions
are required at the wall \cite{SVARD20084805}. 
For example, the no-slip boundary condition is imposed by requiring the three components
of the velocity to be zero and imposing either an 
isothermal or adiabatic condition for temperature.
In this paper, the isothermal boundary condition will be considered.
Note that isothermal implies temperature $T=C_1$ 
where $C_1$ is the given boundary data.
In the following sections on the wall model, the notation $()^*$,
for example, $u^*$ will be used for boundary data. 
For example, the no slip boundary condition is
$u_i^* = 0, i=1,2,3$ and $T^* = C_1$.
The two wall models shown here use either a Neumann or Robin boundary condition.
The other boundary conditions are the no-penetration slip and the isothermal wall.
In a rectangular domain 
$\{(x, y):0 \leq x \leq 1, \ 0 \leq y \leq 1, \ 0 \leq z \leq 1\}$ or $[0, 1]^3$,
with $y=0$ as the wall
the boundary conditions will be implemented as 
$u_i^*=0, i=1,3$, $T^* = C_1$ and a boundary value for 
$\tau^*$ where $\tau$ is the shear stress. 

\subsection{Equilibrium wall stress model}

\begin{figure}[!h]
	 \begin{center}
	   \includegraphics[trim = {0cm 0cm 0cm 0cm}, 
	   					clip, height = 5.5cm, width = 8cm]{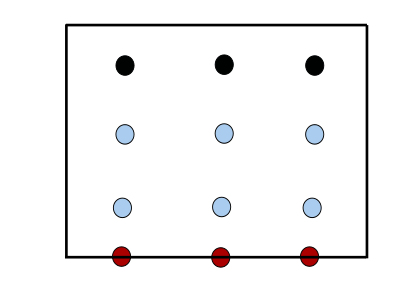}
	 \caption{Illustration of wall modelled element. 
             The solution points are in blue and black, 
             while the red points are the nodes where wall boundary condition is imposed. 
             The velocity $u_{wm}$ used for the wall model 
             is the velocity at the black solution points.}
	 \label{f:wm_scheme}
	 \end{center}
	\end{figure}

The equilibrium wall stress model assumes that 
the flow is fully turbulent and in equilibrium.
Under this assumption the wall stress is determined using 
a log-law like profile \cite{Schlichting2017} for the near wall velocity. 
To present the technique, we first 
define the so called friction velocity as  $u_{\tau} = \sqrt{\tau_w/\rho}$. 
Then the scaled wall normal co-ordinate is defined as 
$y^{+} = yu_{\tau}/\nu$ and the scaled velocity as $u^{+} = u/u_{\tau}$. 
Using these scaled variables, the algebraic \textit{Reichardt function} 
\cite{Reichardt1951} is

		\begin{equation} \label{eq:reic}
			u^+ = \frac{1}{\kappa}\ln(1 + \kappa y^+) + 
			      (C - \frac{1}{\kappa}\ln \kappa )(1 - e^{-\frac{y^+}{11} }
			       - \frac{y^+}{11} e^{-\frac{y^+}{3}} ),
		\end{equation}
			       
\noindent where we choose $C= 4.1$ and $\kappa = 0.38$. 
Given some location $y_{wm}$ away from the wall and 
a velocity $u_{wm}$ at this location,
(\ref{eq:reic}) defines an implicit scalar equation for $\tau_w$.
Given the velocity $\underline{u}_{wm}=[u, v, w]_{wm}$ at $y_{wm}$, 
we simply let $u_{wm} = |\underline{u}_{wm}|$ with $||$ implying magnitude.
The choice of the location $y_{wm}$ is important for the performance of this method.
The convention is to choose a location sufficiently far away from the wall and 
yet at a fraction of the local boundary layer thickness $y_{wm} \sim 0.1\delta$. 
Here, following \cite{Frre2017}, we take the solution point 
farthest away from the wall boundary point 
in the wall adjoining element, see figure \ref{f:wm_scheme}.
Using this velocity, wall stress is determined
such that (\ref{eq:reic}) holds. 
In the case of Gauss Lobatto solution points, 
the wall boundary points are also solution points. 

Given $y_{wm}, u_{wm}$, we solve for the wall stress using (\ref{eq:reic}). 
It is an implicit equation in $\tau_w$ which we solve numerically.
The result has to be imposed as the viscous flux, 
hence we need to convert the scalar value $\tau_w$ to a vector. 
We partition the stress using the velocity vector.
For example, in a domain $[0,1]^3$ the expression for the shear
stress at $y=0$ wall becomes,
    \begin{equation} \label{eq:reic1}
        \begin{aligned}
            \tau^*_{xy} &= {u}_{wm, x}\frac{\tau_w}{|\underline{u}_{wm}|}
        , \
            \tau^*_{yz} &= {u}_{wm, z}\frac{\tau_w}{|\underline{u}_{wm}|}
        \end{aligned}
    \end{equation}
where ${u}_{wm, x}, {u}_{wm, z}$ are the tangential components of the velocity in the 
$x, z$ directions respectively at $y=y_{wm}$
and $\tau^*_{xy}, \tau^*_{yz}$ is the shear stress boundary data.
The process to implement the wall model is summarized below for each wall adjoining element.  

\begin{enumerate}

	\item Loop through boundary points
	
	\item For each boundary point, 
            find the surface normal pointing into the flow. 
            Using this normal find the solution point inside the element 
            that is furthest away from the boundary point along this normal. 
            Take the height of this point from the boundary point as $y_{wm}$.
	
	\item Let the velocity magnitude at that point 
            with height $y_{wm}$ be $u_{wm}$. 
	
	\item Use $\rho, \tau_w$ at the boundary as initial guess for 
            $u_{\tau} = \sqrt{\tau_w/\rho}$ and get initial values for $u^{+}, y^{+}$.
	
	\item Get updated value for $u^{+}, y^{+}$ by substituting previous values in 
            (\ref{eq:reic}). Iterate to convergence for $u_{+}$. 
	
	\item The value of $u_{+}$ thus obtained provides $u_{\tau}$ and thus $\tau_{w}$. 
            Use (\ref{eq:reic1}) to get the shear stress boundary data and 
            impose this as a Neumann boundary condition using 
        simultaneous approximation terms (SAT) \cite{SVARD201417}.

        \item The other boundary conditions used are the no-penetration 
            condition for normal velocity and a fixed temperature isothermal wall.
            Imposition of these boundary conditions are also done using SAT.

\end{enumerate}

\subsection{Dynamic slip wall model}

The dynamic slip wall model was introduced in \cite{Bose2014}.
The boundary condition was derived for the filtered Navier Stokes equations
and reduces to the no-slip condition as the grid is refined. 
It is given by
\begin{equation}  \label{eq:slip_bc}
\begin{aligned}
        u_i - l_p \frac{\partial u_i}{\partial y} = 0, \ i=1,2,3,
\end{aligned}
\end{equation}
at the wall. 
Here $u_i$ indicates the velocity components.
This is a Robin type boundary condition that reduces to the no-slip boundary condition
as $l_p$ goes to zero.
We introduce two changes to the wall model (\ref{eq:slip_bc}).
First, note that (\ref{eq:slip_bc}) allows penetration.   
In a rectangular domain $\{(x, y):0 \leq x \leq 1, \ 0 \leq y \leq 1\}$ or $[0, 1]^2$, 
(\ref{eq:slip_bc}) gives us 
$u_2 - l_p \frac{\partial u_2}{\partial y}$ = 0.
Here $u_2$ is the velocity in $y$ direction.
We modify the boundary condition to directly impose no-penetration,
$u_2 = 0$.
The modified boundary condition is now mass conserving.
Second, we do not use the dynamic procedure from \cite{Bose2014} to obtain $l_p$.
We will treat it as a parameter with a constant positive value
following \cite{CartondeWiart2017}.
We will study the effects of this parameter on the accuracy of the wall model,
in the numerical section.

Note that the Robin boundary condition (\ref{eq:slip_bc})
needs to be converted to boundary data such as (\ref{eq:reic1}).
We do this by using penalization.
To  illustrate this, consider again the domain $[0,1]^3$.
At $y=0$ wall, let the local values for the velocity, shear stress etc.
be $u_1, \tau_{xy}$ .
Using these values the boundary data will be given by
\begin{equation}  \label{eq:slip_bc1}
\begin{aligned}
    \tau^*_{xy} = \tau_{xy} + \sigma (u_1 - l_p \frac{\partial u_1}{\partial y}).
    \\
    \tau^*_{yz} = \tau_{yz} + \sigma (u_3 - l_p \frac{\partial u_3}{\partial y}).
\end{aligned}
\end{equation}
Here $\tau^*_{xy}, \tau^*_{yz}$ is the shear stress boundary data and 
$\tau_{xy}, \tau_{yz}, u_1, u_3, \frac{\partial u_1}{\partial y}, \frac{\partial u_3}{\partial y}$ 
are local shear stress, wall parallel velocity in the $x, z$ directions 
and gradient of wall parallel velocity evaluated at the wall.
$\sigma$ is a parameter, which we will determine based on the kinetic energy stability,
analysed later.
The dynamic slip wall model implementation is summarized below. 

\begin{enumerate}

	\item Loop through boundary points

        \item Choose a constant positive value for the parameter $l_p$.
            Here we will test three cases with $l_p=0.01, 0.1, 0.5$.
	
        \item Using the solution at each boundary point, get  
            $u_1, u_3, \tau_{xy}, \tau_{yz}, \frac{\partial u_1}{\partial y},
            \frac{\partial u_3}{\partial y}$.

        \item Using these values, determine the boundary data 
            from (\ref{eq:slip_bc1})
            and impose using a SAT term.

        \item The other boundary conditions used are the no-penetration 
            condition for normal velocity and a fixed temperature isothermal wall.
            These are also imposed using SAT terms.

\end{enumerate}


\section{Stability for wall model: continuous estimate}

The wall models (\ref{eq:reic1}) and (\ref{eq:slip_bc}) are 
Neumann and Robin type boundary conditions, respectively.
We will now look at the stability properties of the wall models. 
We use the compressible Navier-Stokes (CNS) equations in 2D to derive stability estimates. 
They are given by 
\begin{equation*} \label{eq:cns_again}
\frac{\partial \underline{U}}{\partial t} 
+ 
\nabla \cdot \underline{F}_x^I 
+
\nabla \cdot \underline{G}_y^I
- 
\nabla \cdot \underline{F}_x^V 
-
\nabla \cdot \underline{G}_y^V
= 0
\end{equation*}
with 
\begin{equation*}  \label{eq:cns_a1}
\begin{aligned}
\underline{F}^I &= \left[
\rho u, \rho u^2 + p, \rho u v, u(p + e)
\right]^T
\\
\underline{G}^I &= \left[
\rho v, \rho u v, \rho v^2 + p, v(p + e)
\right]^T
\\
\underline{F}^V &= \left[
0, \tau_{xx}, \tau_{xy}, u\tau_{xx} + v\tau_{xy} + q_x
\right]^T
\\
\underline{G}^V &= \left[
0, \tau_{xy}, \tau_{yy}, u\tau_{xy} + v\tau_{yy} + q_y
\right]^T.
\end{aligned}
\end{equation*}
with 
\begin{equation}  \label{eq:cns_a2}
\begin{aligned}
\tau_{ij} = \mu \left( 
\frac{\partial u_i}{\partial x_j} + \frac{\partial u_j}{\partial x_i}
    -\frac{2}{3}\delta_{ij}\frac{\partial u_k}{\partial x_k}
\right)
, \ 
q_i = \kappa\frac{\partial T}{\partial x_i}.
\end{aligned}
\end{equation}
Using (\ref{eq:cns_a2}), we have the following identities
\begin{equation}  \label{eq:cns_tau_xx}
\begin{aligned}
\frac{1}{\mu}(\tau_{xx} + \frac{1}{2} \tau_{yy}) &&&=  \frac{\partial u}{\partial x} \\
\frac{1}{\mu}(\tau_{yy} + \frac{1}{2} \tau_{xx}) &&&=  \frac{\partial v}{\partial y} \\
\frac{1}{\mu}\tau_{xy} &&&= \frac{\partial u}{\partial x} + \frac{\partial u}{\partial x}.
\end{aligned}
\end{equation}

We will now look at the kinetic energy contribution only from the viscous terms,
and assume that all other terms are bounded. 
The kinetic energy can be derived, by multiplying the continuity equation 
with $-\frac{u^2}{2}$, momentum equation with $u$ and adding them together. 
Since the continuity equation has $0$ viscous terms, 
we will only consider the momentum terms. 
The above procedure gives us
\begin{equation}  \label{eq:ke_wall}
\begin{aligned}
\frac{\partial k}{\partial t}
=
u \frac{\partial \tau_{xx}}{\partial x}
+
u \frac{\partial \tau_{xy}}{\partial y}
+
v \frac{\partial \tau_{xy}}{\partial x}
+
v \frac{\partial \tau_{yy}}{\partial y}
+
conv,
\end{aligned}
\end{equation}
where $conv$ refers to the convective terms.
For the convective terms, 
we will use a kinetic energy conserving scheme.
In the following, we will only consider the viscous terms.

Assume now that we have a rectangular domain. Integrating (\ref{eq:ke_wall}) over the domain yields:
\begin{equation}  \label{eq:ke_wall1}
\begin{aligned}
\int \frac{\partial k}{\partial t} dx dy
=
\int
\left(
u \frac{\partial \tau_{xx}}{\partial x}
+
u \frac{\partial \tau_{xy}}{\partial y}
+
v \frac{\partial \tau_{xy}}{\partial x}
+
v \frac{\partial \tau_{yy}}{\partial y}
\right) dxdy
\end{aligned}.
\end{equation}
Using the chain rule, for example, 
$u \frac{\partial \tau_{xx}}{\partial x} = \frac{\partial u\tau_{xx}}{\partial x}
- \tau_{xx} \frac{\partial u}{\partial x}$ in (\ref{eq:ke_wall1}), results in 
\begin{equation}  \label{eq:ke_wall2}
\begin{aligned}
\int \frac{\partial k}{\partial t} dx dy
=
&\int
\left(
\frac{\partial u \tau_{xx}}{\partial x}
+
\frac{\partial u \tau_{xy}}{\partial y}
+
\frac{\partial v \tau_{xy}}{\partial x}
+
\frac{\partial v \tau_{yy}}{\partial y}
\right) dxdy
\\
-
&\int
\left(
\tau_{xx}\frac{\partial u }{\partial x}
+
\tau_{xy} \frac{\partial u }{\partial y}
+
\tau_{xy}\frac{\partial v }{\partial x}
+
\tau_{yy}\frac{\partial v }{\partial y}
\right) dxdy.
\end{aligned}
\end{equation}
Consider the second term on the right hand side. 
The identity (\ref{eq:cns_tau_xx}) yields
\begin{equation}  \label{eq:ke_diss}
\begin{aligned}
\tau_{xx}\frac{\partial u }{\partial x}
+
\tau_{xy} \frac{\partial u }{\partial y}
+
\tau_{xy}\frac{\partial v }{\partial x}
+
\tau_{yy}\frac{\partial v }{\partial y}
&=
\frac{1}{\mu}\tau_{xx}(\tau_{xx} + \frac{1}{2}\tau_{xx})
+
\frac{1}{\mu}\tau_{yy}(\tau_{xx} + \frac{1}{2}\tau_{yy})
\\
&+
\frac{1}{\mu}\tau_{xy}\tau_{xy}
\\
&=\frac{1}{\mu}(\tau_{xx}^2 + \tau_{xx}\tau_{yy} + \tau_{yy}^2 + \tau_{xy}^2)
.
\end{aligned}
\end{equation}
Now we also have the inequality 
\begin{equation}  \label{eq:iden}
\begin{aligned}
    a^2 + b^2 + ab = \frac{(a+b)^2 + a^2 + b^2}{2}
    \geq 0.
\end{aligned}
\end{equation}
Using (\ref{eq:iden}) in (\ref{eq:ke_diss}) yields
\begin{equation}  \label{eq:ke_diss2}
\begin{aligned}
\tau_{xx}\frac{\partial u }{\partial x}
+
\tau_{xy} \frac{\partial u }{\partial y}
+
\tau_{xy}\frac{\partial v }{\partial x}
+
\tau_{yy}\frac{\partial v }{\partial y}
=
\frac{1}{\mu}(\tau_{xx}^2 + \tau_{xx}\tau_{yy} + \tau_{yy}^2 + \tau_{xy}^2)
\geq 0.
\end{aligned}
\end{equation}
Consequently (\ref{eq:ke_diss2}) and (\ref{eq:ke_wall2}) leads to 
\begin{equation}  \label{eq:ke_wall3}
\begin{aligned}
\int \frac{\partial k}{\partial t} dx dy
\leq
&\int
\left(
\frac{\partial u \tau_{xx}}{\partial x}
+
\frac{\partial u \tau_{xy}}{\partial y}
+
\frac{\partial v \tau_{xy}}{\partial x}
+
\frac{\partial v \tau_{yy}}{\partial y}
\right) dxdy.
\end{aligned}
\end{equation}

Note that (\ref{eq:ke_wall3}) only contains divergence terms. 
Therefore, integrating over the domain leaves only boundary terms. 
We consider a rectangular domain $[0, 1]^2$ 
and investigate the result at the boundary $y = 0$. This gives 
\begin{equation}  \label{eq:ke_wall4}
\begin{aligned}
\int \frac{\partial k}{\partial t} dx dy
&\leq
-
\int
\left(
u^* \tau^*_{xy}
+
v^* \tau^*_{yy}
\right) dx
\end{aligned}
\end{equation}
Using the slip boundary condition 
$u^* = u, v^* = 0 $ 
at the wall leads to
\begin{equation}  \label{eq:ke_wall_star}
\begin{aligned}
\int \frac{\partial k}{\partial t} dx dy
&\leq
-\int
u^* \tau^*_{xy}
dx
.
\end{aligned}
\end{equation}
Expanding (\ref{eq:ke_wall_star}) with $u^* = u$ and 
$\tau^*_{xy} = \mu(\frac{\partial u}{\partial y} + \frac{\partial v}{\partial x})$
\begin{equation}  \label{eq:ke_wall4_2}
\begin{aligned}
\int \frac{\partial k}{\partial t} dx dy
&\leq
-\int
u \tau^*_{xy}
dx
=
-
\int
u \mu(\frac{\partial u}{\partial y} + \frac{\partial v}{\partial x})
dx
=
-
\int
\mu u \frac{\partial u}{\partial y}
dx
.
\end{aligned}
\end{equation}
Therefore, to bound kinetic energy, 
$\frac{\partial u}{\partial y}$ and $u$ must have opposite signs.

\subsection{Stability for equilibrium wall stress model}

If we use the equilibrium wall model, the boundary data is given by 
\begin{equation*} 
    {\tau}^*_{xy} = {u}_{wm, x}\frac{\tau_w}{|u_{wm}|}
\end{equation*}
%
%
%
%
This would lead to 
\begin{equation}  \label{eq:ke_wall5}
\begin{aligned}
\int \frac{\partial k}{\partial t} dx dy
&\leq
-\int
u \tau^*_{xy}
dx
=
-
\int
u {u}_{wm, x}\frac{\tau_w}{|u_{wm}|}
dx
.
\end{aligned}
\end{equation}
Since, $u, u_{wm, x}$ in general do not have the same sign, 
the wall model is not kinetic energy stable. 
This is particularly serious for highly sheared flows.

\subsection{Stability for dynamic slip wall model}

For the dynamic slip wall model, the stability estimate 
using (\ref{eq:ke_wall4_2}) becomes
\begin{equation}  \label{eq:ke_dyn}
\begin{aligned}
\int \frac{\partial k}{\partial t} dx dy
&\leq
-
\int
    \frac{\mu}{l_p}u^2 
dx
.
\end{aligned}
\end{equation}
The dynamic slip wall model is therefore kinetic energy stable,
since the length $l_p > 0$.
The implementation (\ref{eq:slip_bc1})
in the continuous setting using (\ref{eq:ke_wall_star})
becomes
\begin{equation}  \label{eq:ke_wall6}
\begin{aligned}
\int \frac{\partial k}{\partial t} dx dy
&\leq
    -\int
u \tau^*_{xy}
dx
=
-
\int
    u
    \left(\mu \frac{\partial u}{\partial y} + \sigma (u - l_p \frac{\partial u}{\partial y})
    \right)
dx
.
\end{aligned}
\end{equation}
Specifying the penalty parameter $\sigma$ as 
$
\sigma = \frac{\mu}{l_p}
$
in (\ref{eq:ke_wall6})
results in the stability estimate
\begin{equation}  \label{eq:ke_wall7}
\begin{aligned}
\int \frac{\partial k}{\partial t} dx dy
&\leq
-
\int
    \frac{\mu}{l_p}
  u^2, 
dx
.
\end{aligned}
\end{equation}
%
%
%
%
%
\section{Numerical Method}\label{sec:num}

In this section we review 
our specific discontinuous Galerkin method, 
the associated SBP property and how it leads to energy stability.

\subsection{Notation} 

An underlined lower case letter denotes a vector quantity, e.g. $\underline{u}$.  
A doubly underlined lower case letter denotes a vector imposed on the diagonal of a matrix, i.e. $\doubleunderline{u} = diag(\underline{u})$.  
A doubly underlined upper case letter denotes any other matrix, e.g. $\doubleunderline{U}$. 
A variable with no underline denotes a scalar quantity.  
Quantities marked as $()^*$, for example, $u^*$ denotes boundary data.
%


\subsection{DG and Summation-By-Parts} \label{sec:DG}

Consider the linear advection equation. 
\begin{equation} \label{eq:adv}
\begin{aligned}
\frac{\partial u}{\partial t} + \frac{\partial u}{\partial x} = 0.
\end{aligned}
\end{equation}
If we multiply (\ref{eq:adv}) by u and integrate over the domain, we get
\begin{equation} \label{eq:ibp1}
\begin{aligned}
\int_{\Omega} 2uu_t dx = 
\frac{d }{d t} \int_{\Omega} u^2 dx 
    &= -\int_{\Omega} 2u\frac{\partial u}{\partial x}  dx 
    &= -u^2 |_{d\Omega} 
\end{aligned}
\end{equation}
Therefore, the energy integral over the domain only depends on the values of 
$u^2$ at the boundaries. 

Following \cite{Gassner2013, RANOCHA2016299}, 
we can discretize this using the strong form of DG at Gauss Lobatto nodes 
also known as the discontinuous Galerkin collocation spectral element method (DGSEM)
as
\begin{equation} \label{eq:linear_sbp}
\frac{\partial \underline{u}}{\partial t} + \doubleunderline{D} \ \underline{u} + 
\doubleunderline{M}^{-1}  \doubleunderline{R}^T  \doubleunderline{B}
(\underline{u}^* - \doubleunderline{R} \  \underline{u})= 0.
\end{equation}
The operators $\doubleunderline{D}, \doubleunderline{R}$ 
represent the derivative and restriction matrices respectively. 
The restriction matrix can be defined as 
$$
\doubleunderline{R} \ \underline{u}
				=
				 \begin{bmatrix}
				     u_L \\
				     u_R
				  \end{bmatrix}
$$
where $u_L, u_R$ are the left and right values of $u$ at the edge of an element and  
$u^{*}$ is the numerical flux (eg. Rusanov flux \cite{Toro2009}).

Summation-By-Parts (SBP) is a discrete analogue of integration by parts 
that allows extension of continuous stability estimates to the discrete one's 
see \cite{SVARD201417}. 
First, we define a derivative and quadrature matrix 
$\doubleunderline{D}, \doubleunderline{M}$ such that
$$
\doubleunderline{D} \underline{u} \approx \partial_x u,
\
\underline{u}^T \doubleunderline{M} \underline{v} \approx
\int u v dx. 
$$
The derivative matrix for DGSEM can simply be obtained by 
using derivatives of Lagrange polynomials at Gauss Lobatto nodes
while the quadrature matrix is 
$\doubleunderline{M} = diag(\omega_0, \omega_1, \dots, \omega_P)$, 
where $\omega_j, \, j=0,\dots,P$ denotes quadrature weights.
Next define the boundary matrix as 
$\doubleunderline{B} = diag(-1, 1)$.
The matrix $\doubleunderline{R}$ then satisfies  
$$
\doubleunderline{R}^T \doubleunderline{B} \ \doubleunderline{R}= 
diag(-1, 0,.., 0, 1).
$$
With these definitions, the SBP operator satisfies the following property:
\begin{equation} \label{eq:sbp1}
\doubleunderline{M} \ \doubleunderline{D} + \doubleunderline{D}^T  \doubleunderline{M} = 
\doubleunderline{R}^T \doubleunderline{B} \ \doubleunderline{R}.
\end{equation}
The symmetric positive definite matrix $\doubleunderline{M}$ defines a norm 
$||\underline{u}||^2_M = \underline{u}^T \doubleunderline{M} \ \underline{u}$. 
Integration by parts can be mimicked using these operators since:
\begin{equation*} \label{eq:ibp}
\begin{aligned}
\underline{u}^T \doubleunderline{M} \ \doubleunderline{D} \ \underline{v} +
\underline{v}^T \doubleunderline{M} \ \doubleunderline{D} \underline{u}
=
(\doubleunderline{R} \ \underline{u})^T \doubleunderline{B} \ (\doubleunderline{R} \ \underline{v})
&\approx \int_{-1}^{1} u\partial_x v  + \int_{-1}^{1} v\partial_x u
= uv|_{-1}^{1}
\end{aligned}
\end{equation*}
Next, we consider a single element and 
for a moment ignore the simultaneous approximation term (SAT). 
Multiply (\ref{eq:linear_sbp}) on the left by 
$2\doubleunderline{u}^T\doubleunderline{M}$
which by using (\ref{eq:sbp1}) leads to
\begin{equation} \label{eq:sbp_adv4}
\begin{aligned}
2\underline{u}^T \doubleunderline{M} \
\frac{d }{d t}  \underline{u}
=
\frac{d }{d t}  \underline{u^2}
&=
-
(
\underline{u}^T  \doubleunderline{M} \ 
\doubleunderline{D} \ \underline{u}
+
\underline{u}^T \  \doubleunderline{D}^T  
\doubleunderline{M} \ \underline{u}
)
\\
&=
-
\underline{u}^T
\doubleunderline{R}^T  \doubleunderline{B} \ \doubleunderline{R} \
\underline{u}
= u_L^2 - u_R^2.
\end{aligned}
\end{equation}
Clearly, (\ref{eq:sbp_adv4}) is the exact discrete equivalent of (\ref{eq:ibp1}).
Suitable boundary conditions can be imposed using the SAT term.
Therefore, we discretely mimicked the continuous stability estimate using the SBP property. 

The solution of the compressible Navier Stokes(CNS) equations
requires discretization of viscous terms. 
This will be illustrated using the advection diffusion equation which 
can be written as
\begin{equation*} \label{eq:adv-diff}
\begin{aligned}
\frac{\partial u}{\partial t} + \frac{\partial u}{\partial x} 
    = \epsilon \frac{\partial^2 u}{\partial x^2},
\end{aligned}
\end{equation*}
or  
\begin{equation} \label{eq:adv-diff1}
\begin{aligned}
\frac{\partial u}{\partial t} + \frac{\partial u}{\partial x} 
    &= \epsilon \frac{\partial q}{\partial x} ,
    \\
    q &=  { {\partial u} \over {\partial x} }.
\end{aligned}
\end{equation}
Equation (\ref{eq:adv-diff1}) can be discretized as
\begin{equation} \label{eq:linear_advdiff}
\begin{aligned}
\frac{\partial \underline{u}}{\partial t} + \doubleunderline{D} \ \underline{u} + 
\doubleunderline{M}^{-1}  \doubleunderline{R}^T  \doubleunderline{B}
(\underline{u}^* - \doubleunderline{R} \  \underline{u})
    &= \epsilon \big( \doubleunderline{D} \ \underline{q} + 
\doubleunderline{M}^{-1}  \doubleunderline{R}^T  \doubleunderline{B}
    (\underline{q}^* - \doubleunderline{R} \  \underline{q}) \big)
    \\
    \underline{q} &= \doubleunderline{D} \ \underline{u} + 
\doubleunderline{M}^{-1}  \doubleunderline{R}^T  \doubleunderline{B}
(\underline{u}^* - \doubleunderline{R} \  \underline{u})
\end{aligned}
\end{equation}
The advection diffusion equation can have 3 types of boundary conditions.
They are Dirichlet $u=g_1$, Neumann $u_x=g_2$ or Robin $\alpha u - \beta u_x =g_3$
where $g_1, g_2, g_3$ is boundary data. 
Dirichlet boundary condition is imposed using $\underline{u}^*=g_1$ and 
$\underline{q}^* = \doubleunderline{R} \  \underline{q}$.
Neumann boundary condition is imposed using $\underline{q}^*=g_2$ and 
$\underline{u}^* = \doubleunderline{R} \  \underline{u}$.
Robin boundary condition is imposed using 
$\underline{u}^* = \doubleunderline{R} \  \underline{u}$ and
$\underline{q}^* = \doubleunderline{R} \  \underline{q} + 
\sigma( \alpha \doubleunderline{R} \  \underline{u} 
- \beta  \doubleunderline{R} \  \underline{q} - g_3)$ .
Here $\sigma$ is a suitably chosen penalty parameter 
such as the one chosen in (\ref{eq:ke_wall7}).

In the interest of keeping the details to a minimum, we end this section by mentioning that
we will use the Kennedy and Gruber split form \cite{KENNEDY2008}
which is kinetic energy conserving \cite{GASSNER2016},
in the Navier Stokes calculations. 


\section{Stability for wall model: discrete estimate} \label{sec:dis_sbp}

To get discrete estimates in $d$ dimensions, 
we will use the tensor product formulation. 
We use the notation
$\doubleunderline{m}, \doubleunderline{d}, \doubleunderline{r}, \doubleunderline{b}$ 
to denote the 1 dimensional mass, derivative, restriction and boundary matrices.
Next we need the Kronecker product. 
Consider two matrices 
$\doubleunderline{a} \in \mathbb{R}^{k \times l}, \ \doubleunderline{b} \in \mathbb{R}^{m \times n}$
Then we define the Kronecker product as
\begin{equation} \label{eq:kron}
\doubleunderline{a}
\otimes
\doubleunderline{b}
:=
\left(
\begin{aligned}
&a_{11} \doubleunderline{b} \quad a_{12} \doubleunderline{b}
\quad &&\dots \quad &&&a_{1l} \doubleunderline{b}
\\
&\vdots &&\ddots &&&\vdots
\\
&a_{k1} \doubleunderline{b} \quad a_{k2} \doubleunderline{b}
\quad &&\dots \quad &&&a_{kl} \doubleunderline{b}
\end{aligned}
\right)
\end{equation}
Define the required two dimensional matrices as
\begin{equation} \label{eq:kron_mat}
\begin{aligned}
\doubleunderline{M} &=
\doubleunderline{m}
\otimes
\doubleunderline{m}, \quad
\doubleunderline{D_1} =
\doubleunderline{I}
\otimes
\doubleunderline{d}, \quad
\doubleunderline{D_2} =
\doubleunderline{d}
\otimes
\doubleunderline{I}
\\
\doubleunderline{R_1} &=
\doubleunderline{I}
\otimes
\doubleunderline{r}, \quad
\doubleunderline{R_2} =
\doubleunderline{r}
\otimes
\doubleunderline{I}, \quad
\doubleunderline{B_1} =
\doubleunderline{I}
\otimes
\doubleunderline{b} \quad
\doubleunderline{B_2} =
\doubleunderline{b}
\otimes
\doubleunderline{I}
\end{aligned}
\end{equation}
In this case the SBP formulation (\ref{eq:sbp1}) becomes (see \cite{RANOCHA201713})
\begin{equation} \label{eq:sbp_d}
\begin{aligned}
\doubleunderline{M} \ \doubleunderline{D_i} + \doubleunderline{D_i}^T  \doubleunderline{M} = 
\doubleunderline{R_i}^T \doubleunderline{B_i} \ \doubleunderline{R_i}
\end{aligned}
\end{equation}
We now have the derivative operators $\doubleunderline{D_i}$ acting in the $i^{th}, i=1,..,d$ direction. 
These operators satisfy the multi-dimensional SBP conditions 
(see \cite{ranocha_thesis}). 
Next, we look at the discretization of the viscous terms. 
The stress is discretized as 
\begin{equation} \label{eq:tau}
\begin{aligned}
\underline{ \tau_{xx} }
&=
\frac{4}{3} \mu \doubleunderline{D_1} \ \underline{ u }  
-
\frac{2}{3} \mu 
\doubleunderline{D_2} \ \underline{ v }
\\
&+
\frac{4}{3} \mu 
\doubleunderline{M}^{-1}  \doubleunderline{R_1}^T  \doubleunderline{B_1}
(\underline{u}^* - \doubleunderline{R_1} \  \underline{u})
-\frac{2}{3} \mu 
\doubleunderline{M}^{-1}  \doubleunderline{R_2}^T  \doubleunderline{B_2}
(\underline{v}^* - \doubleunderline{R_2} \  \underline{v})
\\
\underline{ \tau_{yy} }
&=
\frac{4}{3} \mu \doubleunderline{D_2} \ \underline{ v }  
-
\frac{2}{3} \mu 
\doubleunderline{D_1} \ \underline{ u }
\\
&+
\frac{4}{3} \mu 
\doubleunderline{M}^{-1}  \doubleunderline{R_2}^T  \doubleunderline{B_2}
(\underline{v}^* - \doubleunderline{R_2} \  \underline{v})
-\frac{2}{3} \mu 
\doubleunderline{M}^{-1}  \doubleunderline{R_1}^T  \doubleunderline{B_1}
(\underline{u}^* - \doubleunderline{R_1} \  \underline{u})
\\
\underline{ \tau_{xy} }
&=
\mu 
\left(
\doubleunderline{D_2} \ \underline{ u } + \doubleunderline{D_1} \ \underline{ v }
\right)
\\
&+
\mu 
\doubleunderline{M}^{-1}  \doubleunderline{R_1}^T  \doubleunderline{B_1}
(\underline{v}^* - \doubleunderline{R_1} \  \underline{v})
+
\mu 
\doubleunderline{M}^{-1}  \doubleunderline{R_2}^T  \doubleunderline{B_2}
(\underline{u}^* - \doubleunderline{R_2} \  \underline{u})
\end{aligned}
\end{equation}
where $u^*, v^*$ are boundary data. 
To impose the no-penetration boundary condition, we use
$\underline{u}^* = \doubleunderline{R_2} \  \underline{u},
\ \underline{v}^* = 0$.

The temperature wall boundary condition is imposed 
in the energy equation which does not contribute to the
kinetic energy bound.
The isothermal wall boundary condition is imposed as 
$T^*=C$ where $C$ is the desired constant value.
From (\ref{eq:tau}) it is obvious that 
\begin{equation} \label{eq:tau1}
\begin{aligned}
\underline{ \tau_{xx} }
+
\frac{1}{2}
\underline{ \tau_{yy} }
&=
\mu \doubleunderline{D_1} \ \underline{ u }  
+
\mu 
\doubleunderline{M}^{-1}  \doubleunderline{R_1}^T  \doubleunderline{B_1}
(\underline{u}^* - \doubleunderline{R_1} \  \underline{u})
\\
\underline{ \tau_{yy} }
+
\frac{1}{2}
\underline{ \tau_{xx} }
&=
\mu \doubleunderline{D_2} \ \underline{ v }  
+
\mu 
\doubleunderline{M}^{-1}  \doubleunderline{R_2}^T  \doubleunderline{B_2}
(\underline{v}^* - \doubleunderline{R_2} \  \underline{v})
\end{aligned}
\end{equation}
Stress gradients are given by
\begin{equation} \label{eq:tau2}
\begin{aligned}
\underline{ \tau_{xx, x} }
&=
\doubleunderline{D_1} \ \underline{ \tau_{xx} }  
+
\doubleunderline{M}^{-1}  \doubleunderline{R_1}^T  \doubleunderline{B_1}
(
\underline{ \tau_{xx}^{*} } - \doubleunderline{R_1} \ \underline{\tau_{xx}}
)
\\
\underline{ \tau_{xy, x} }
&=
\doubleunderline{D_1} \ \underline{ \tau_{xy} }  
+
\doubleunderline{M}^{-1}  \doubleunderline{R_1}^T  \doubleunderline{B_1}
(
\underline{ \tau_{xy}^{*} } - \doubleunderline{R_1} \ \underline{\tau_{xy}}
)
\\
\underline{ \tau_{xy, y} }
&=
\doubleunderline{D_2} \ \underline{ \tau_{xy} }  
+
\doubleunderline{M}^{-1}  \doubleunderline{R_2}^T  \doubleunderline{B_2}
(
\underline{ \tau_{xy}^{*} }- \doubleunderline{R_2} \ \underline{\tau_{xy}}
)
\\
\underline{ \tau_{yy, y} }
&=
\doubleunderline{D_2} \ \underline{ \tau_{yy} }  
+
\doubleunderline{M}^{-1}  \doubleunderline{R_2}^T  \doubleunderline{B_2}
(
\underline{ \tau_{yy}^{*} } - \doubleunderline{R_2} \ \underline{\tau_{yy}}
).
\end{aligned}
\end{equation}
where  
$(\cdot)_{,x}$ denotes the derivative with respect to $x$. 
We can impose the boundary data (\ref{eq:reic1}), (\ref{eq:slip_bc1})
using $\underline{ \tau_{xy}^{*} }$ in (\ref{eq:tau2}).
For $\underline{ \tau_{xx}^{*} }, \underline{ \tau_{yy}^{*} }$
we use
$
\underline{ \tau_{xx}^{*} } = \doubleunderline{R_1} \ \underline{\tau_{xx}},
\underline{ \tau_{yy}^{*} } = \doubleunderline{R_2} \ \underline{\tau_{yy}}.
$
Recall that we aim for the discrete version of kinetic energy (\ref{eq:ke_wall1}). 
\begin{equation} \label{eq:dis_ke}
\begin{aligned}
    \frac{d }{d t} \underline{1}^T  \doubleunderline{M} \ \underline{k}
=
\underline{u}^T \doubleunderline{M} \underline{ \tau_{xx, x} }
+
\underline{u}^T \doubleunderline{M} \underline{ \tau_{xy, y} }
+
\underline{v}^T \doubleunderline{M} \underline{ \tau_{xy, x} }
+
\underline{v}^T \doubleunderline{M} \underline{ \tau_{yy, y} }.
\end{aligned}
\end{equation}
Using the SBP property and velocity boundary data, we get
\begin{equation} \label{eq:ke_tau_y}
\begin{aligned}
\frac{d }{d t} \underline{1}^T  \doubleunderline{M} \ \underline{k}
&\leq
\underline{ \tau_{xy} }^T
\doubleunderline{R_2}^T  \doubleunderline{B_2} \ \doubleunderline{R_2}\ \underline{u}
+
\underline{u}^T  \doubleunderline{R_2}^T  \doubleunderline{B_2}
(
\underline{ \tau_{xy}^{*} }- \doubleunderline{R_2} \ \underline{\tau_{xy}}
)
\\
&=
\underline{u}^T  \doubleunderline{R_2}^T  \doubleunderline{B_2} \
\underline{ \tau_{xy}^{*} }.
\end{aligned}
\end{equation}
%
%
%
which is the discrete equivalent of (\ref{eq:ke_wall_star}). 
The full derivation is presented in \ref{ap:ke} 
and the conclusion is that the discrete kinetic energy stability 
estimates mimic the continuous kinetic energy estimates.


\section{Results} \label{sec:res}

In this section, we will present numerical results.
The properties of the underlying DGSEM and split form is shown in
\cite{Singh2020}.
First, the isentropic vortex test case is used to show the
order of convergence of the numerical method.
Next, we use the turbulent channel flow 
to compare the accuracy of the wall models. 
We then consider a simple 2D airfoil to show that 
kinetic energy instability of the wall model can lead to solution blow-up. 


\subsection{Isentropic vortex}

The 2D isentropic vortex problem is an exact solution for the compressible Euler equations. 
It can be used to test the order of accuracy of a numerical method.
There are various initializations reported for this problem, 
such as shown in \cite{nasa_fr}.  
In this study, the domain $\left[ -5, 5\right]^2$ is considered with the following initial condition:
\begin{equation*} \label{eq:ivortex}
\begin{aligned}
\rho &= \left[ 1 - \frac{\beta^2(\gamma - 1)}{8\gamma\pi^2} \exp(1 - r^2) \right] \\
u &= M -\frac{\beta (y - y_c)}{2\pi} \exp(\frac{1 - r^2)}{2}  \\
v &= \ \ \ \ \  \ \  \frac{\beta (y - y_c)}{2\pi} \exp(\frac{1 - r^2}{2})
\end{aligned}
\end{equation*}
where $r = \sqrt{(x - x_c)^2 + (y - y_c)^2}$.  
Also, $p = \rho^{\gamma}, \beta = 5, \gamma = 1.4$, 
where $\beta$ determines vortex strength.  
Further, $(x_c, y_c) = (0, 0), M = 0.5$, 
and periodic boundary conditions are imposed. 
All simulations are performed at $CFL = 0.1$. 


\begin{table}[!htbp]
	\centering
	\caption{$L^1$ error and order of convergence for the isentropic vortex.}	
		\begin{tabular}{lcc c cc c cc}
		\hline
		 & \multicolumn{2}{c}{$P=3$} && \multicolumn{2}{c}{$P=4$} 
		 \\ 
  		\cline{2-3} \cline{5-6} 
		N     & $L^1$ error   & OC   && $L^1$ error   & OC   
		\\
		\hline
		6  & 1.03E-2  &   -  && 
    	           1.92E-3  &   -  \\
		12  & 8.19E-4 & 3.65 &&
		1.04E-4 & 4.21\\
		24  & 7.49E-5 & 3.45 &&
		5.32E-6 & 4.29\\
		48  & 7.34E-6 & 3.35 &&
		2.48E-7 & 4.42\\
		96  & 5.00E-7 & 3.88 &&
		1.26E-8 & 4.30\\
		\hline
		\end{tabular}
	\label{tab:eoc_p3}
\end{table}

Tables \ref{tab:eoc_p3} shows the error and order of convergence (OC) 
obtained using $N^2$ elements, after one flow through. 
Two polynomial order, $P=3, 4$ are shown.
The accuracy is higher than order $P$ but slightly less than 
the expected order of $P+1$.
This is due to the fact that Gauss Lobatto points
generally leads to lower accuracy than Gauss Legendre points \cite{Kopriva2010}. 

\subsection{Turbulent channel flow} 

One of the objectives in developing high order methods for unstructured grids 
is to enable simulation of wall bounded turbulent flows. 
The simplest possible test case for this is turbulent channel flow \cite{Kim1987}.
This flow has a periodic domain 
in the streamwise and spanwise directions with walls at the top and bottom.
The flow is driven by a forcing term to get a constant mass flow rate. 
The test case is characterized by the parameter $Re_{\tau}$ given by 
$$
Re_{\tau} = \frac{u_\tau L}{\nu}, \ u_\tau = \sqrt{\frac{\tau_w}{\rho}}.
$$
%
%
	 \begin{figure}[!h]
	 \begin{center}
	   \includegraphics[trim = {0cm 0cm 0cm 0cm}, 
	   					clip, height = 6.5cm, width = 14cm]{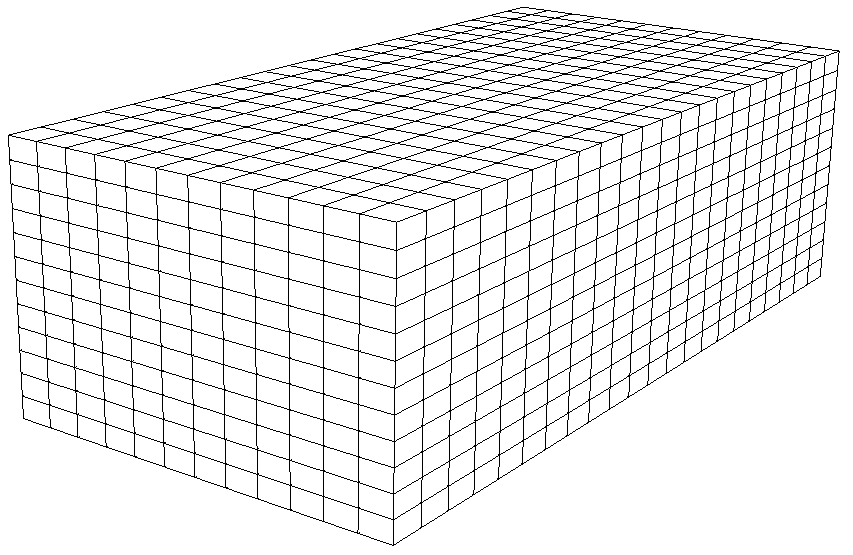}
	 \caption{Uniform mesh for channel flow.}
	 \label{f:tcf_mesh1}
	 \end{center}
	\end{figure}
	 \begin{figure}[hbt!]
	 \begin{center}
	   \includegraphics[trim = {0cm 0cm 0cm 0cm}, 
 					clip, width = 12cm]{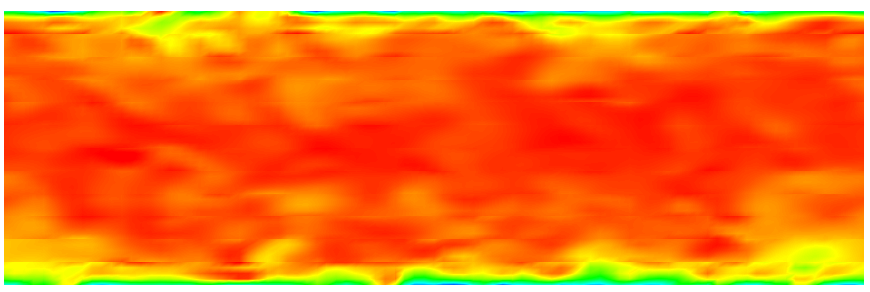}
                                        \caption{Instantaneous snapshot showing horizontal velocity contours for flow with equilibrium wall model}
	 \label{f:wm_flow}
	 \end{center}
	\end{figure}
DNS studies have been done for this case at 
$Re_{\tau} = 5200$ \cite{Kim1987, moser_2015}. 
Studies have also been done using spectral type methods 
such as flux reconstruction (FR) and DG \cite{Vermeire2016, Lodato2012}.
We use a domain length $2\pi$ in the streamwise direction and 
$\pi$ in the spanwise direction.  
The channel height is $H=2$. 
It is discretized into $22 \times 12 \times 12$ uniform elements. 
Figure \ref{f:tcf_mesh1} shows the mesh. 
The computations were compared to DNS data from \cite{moser_2015}.

The initialization and forcing is similar to \cite{Lodato2012}, 
with the streamwise velocity initialized using the profile,
$$
u  = \frac{9}{5} u_0 (1 - (y - 1)^2)^2. 
$$
The vertical velocity in initialized using
$$
v  = 0.1u_0
\exp
\left[
-\left(\frac{x-\pi}{2\pi}\right)^2
\right]
\exp
\left[
-\left(\frac{y}{2}\right)^2
\right]
\cos(4z).
$$
	 \begin{figure}[hbt!]
	 \begin{center}
	   \includegraphics[trim = {0cm 0cm 0cm 0cm}, 
 					clip, height = 12cm, width = 14cm]{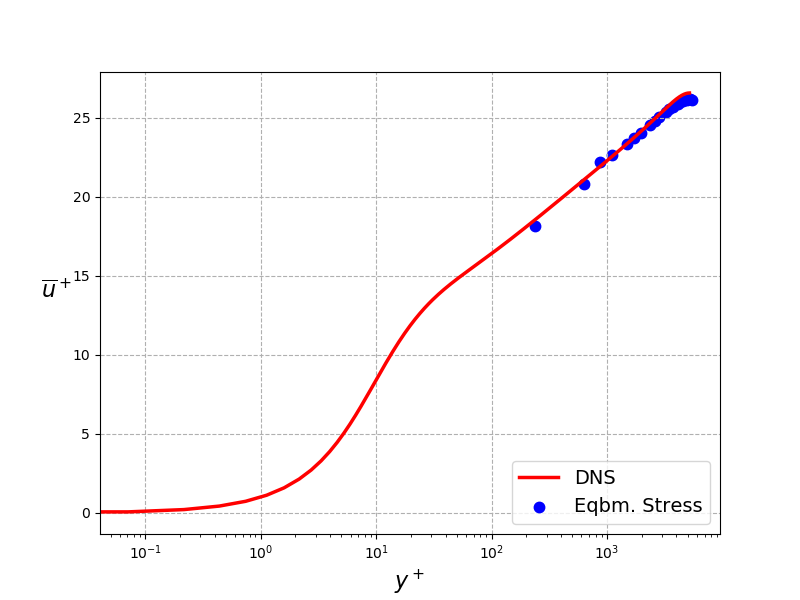}
                                        \caption{Time averaged mean profiles of streamwise velocity at $Re_\tau$ = 5200 for wall modelled ILES (WMILES) at P=3 using equilibrium stress wall model compared with DNS from \cite{moser_2015}.}
	 \label{f:tcf_5200_wm_1}
	 \end{center}
	\end{figure}
	 \begin{figure}[hbt!]
	 \begin{center}
	   \includegraphics[trim = {0cm 0cm 0cm 0cm}, 
 					clip, height = 12cm, width = 14cm]{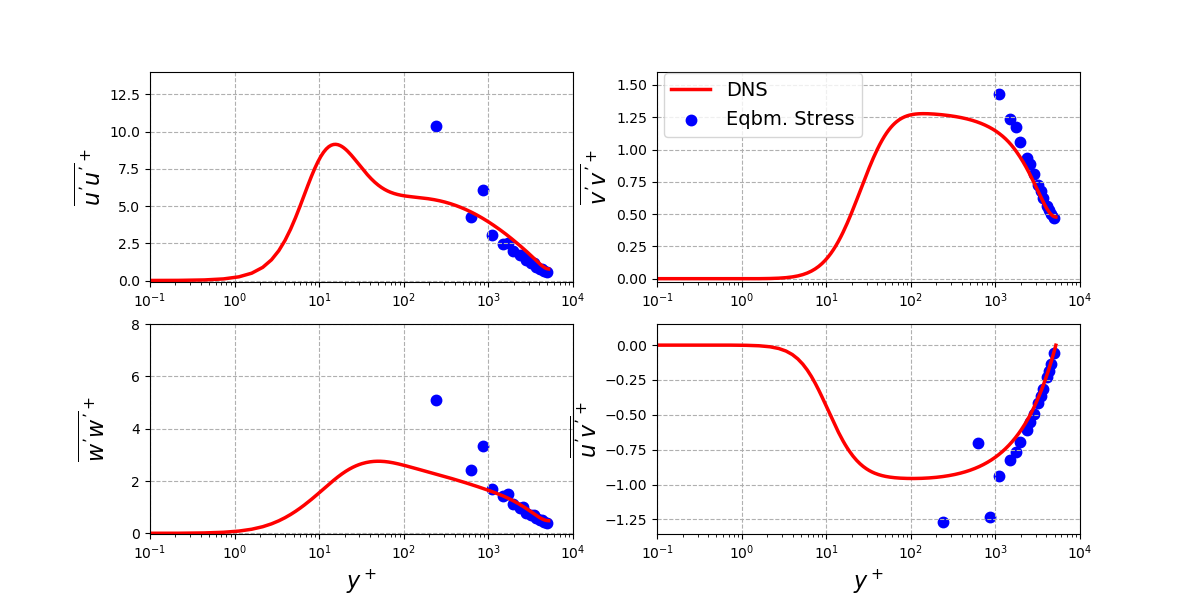}
                                        \caption{Time averaged mean profiles of second moments at $Re_\tau$ = 5200 for wall modelled ILES (WMILES) at P=3 using equilibrium stress wall model compared with DNS from \cite{moser_2015}.}
	 \label{f:tcf_5200_var_tau}
	 \end{center}
	\end{figure}
\begin{figure}[hbt!]
    \begin{center}
        \includegraphics[trim = {0cm 0cm 0cm 0cm}, 
          clip, height = 10cm, width = 14cm]{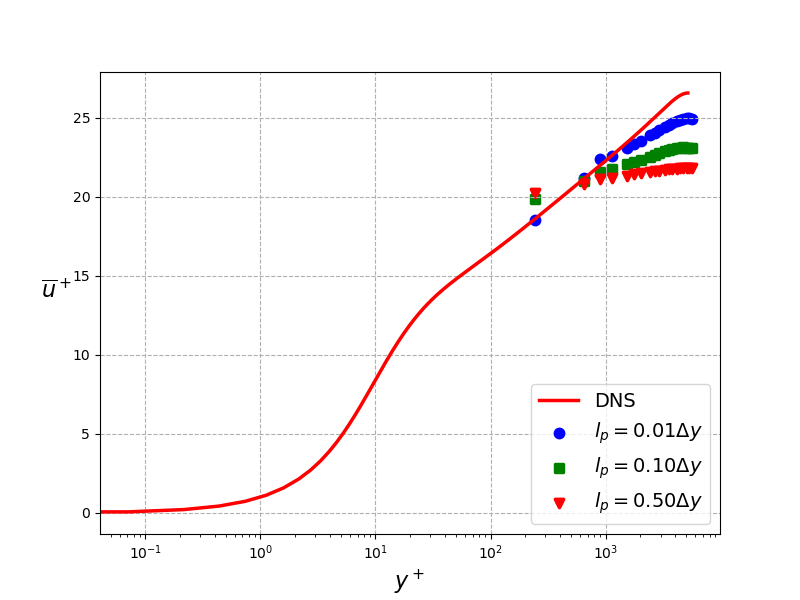}
        \caption{Time averaged mean profiles of streamwise velocity at $Re_\tau$ = 5200 
        for wall modelled ILES (WMILES) using dynamic slip wall model compared with DNS from \cite{moser_2015}.}
        \label{f:tcf_5200_wm_2}
    \end{center}
\end{figure}
\begin{figure}[hbt!]
    \begin{center}
        \includegraphics[trim = {0cm 0cm 0cm 0cm}, 
          clip, height = 10cm, width = 14cm]{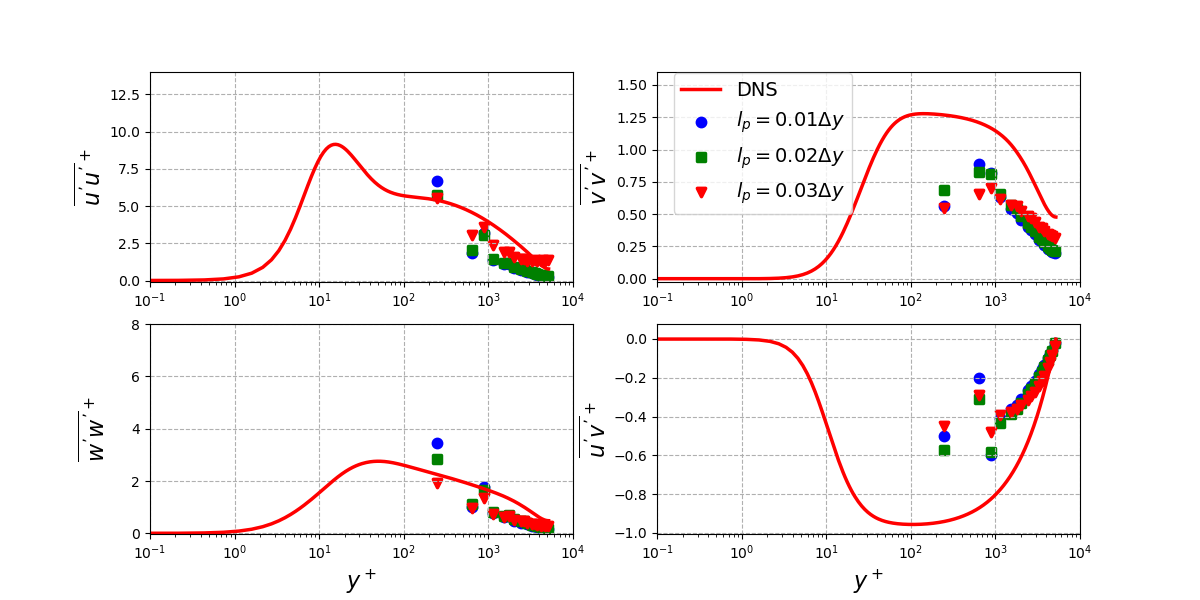}
        \caption{Time averaged mean profiles of second moments at $Re_\tau$ = 5200 
        for wall modelled ILES (WMILES) using dynamic slip wall model compared with DNS from \cite{moser_2015}.}
        \label{f:tcf_5200_var_lp}
    \end{center}
\end{figure}
The initial density is taken as $\rho = 1$ and 
the pressure chosen such that the Mach number for velocity $u_0$ is $M = 0.2$. 
To this uniform profile, perturbations are added to trigger transition.
They are of the form
$$
u_\epsilon = 0.8\sin(\lambda \frac{\pi y}{2})\sin(\lambda \frac{\pi z}{2})
$$
where $\lambda$ is a suitable parameter, eg. $\lambda = 20, 30, 40$. 
To enable constant mass flow rate, 
a source term is added to the right hand side of (\ref{eq:cns_again})
of the form
\begin{equation} \label{eq:tcf_f}
\underline{S} = (0, s_1, 0, 0, s_2)^T, \ 
s_1 = \frac{F_w}{V} + \frac{0.3}{dt}(m_{Re} - 2 m + m_0), \ 
s_2 = u_b s_1
\end{equation}
Here $F_w, m_0, m, m_{Re}$ are the shear at the wall, mass flow at the previous time step, mass flow at the current time step and mass flow required for the prescribed $Re_{\tau}$ respectively.  
The first term in $s_1$ 
balances the total shear stress generated at the wall, 
while the second term accelerates convergence. 
The term $s_2$ is there simply to be energy consistent with the forcing 
where $u_b$ is the average streamwise velocity over the domain.
Figure \ref{f:wm_flow} shows horizontal velocity contours for this flow
with the equilibrium stress wall model.
The flow is clearly highly turbulent.
In addition, the boundary layer is very thin as expected for 
a very high Reynolds' number flow.

\subsubsection{The equilibrium stress model}

Figure \ref{f:tcf_5200_wm_1} shows the mean velocity profile averaged over time for $Re_\tau$ = 5200 with the equilibrium stress wall model. 
Notice first that the first two points are slightly off from the DNS curve. 
Recall that in the wall model, the tangential velocity at the top solution point in the wall adjacent element is taken as the input for the shear stress imposed at the wall. 
Therefore, the top point is assumed to be the first point that is within the resolved region of the flow.
This indicates that the first 3 points probably do not match the DNS data. 
We notice that from the third point onwards, our wall modeled data compares well with the DNS data. 
This is despite the fact that our first solution point is at $y^{+} \sim 200$ which is in the log-layer. 
Therefore, with the wall model we see that the statistics of the first moment are well captured. 
Figure \ref{f:tcf_5200_var_tau} shows the second moments.
Similar to the mean velocity, the simulations depart near the boundary
but away from the wall, they compare well with DNS data.

\subsubsection{Dynamic slip wall model}

Next we use the dynamic slip wall model and compare it to DNS data.
We use $l_p = C \Delta y$ where $C, \Delta y$ 
denotes a constant and the height of the first element respectively.
We test three cases with $C=0.01, 0.10, 0.50$. 
Figure \ref{f:tcf_5200_wm_2} compares the mean velocity profile with DNS data.
We note that there is a large mismatch between wall modeled results
and DNS data.
However, as the values of $C$ is being reduced, the results gradually improve 
but the mismatch persists.
Figure \ref{f:tcf_5200_var_lp} shows the second moments,
which also deviates from DNS data.
This indicates that the choice of a constant $l_p$ is too simplistic.

%


\subsection{Stability issues for sheared flows}

	 \begin{figure}[hbt!]
	 \begin{center}
	   \includegraphics[trim = {0cm 0cm 0cm 0cm}, 
	   					clip, height = 6cm, width = 14cm]{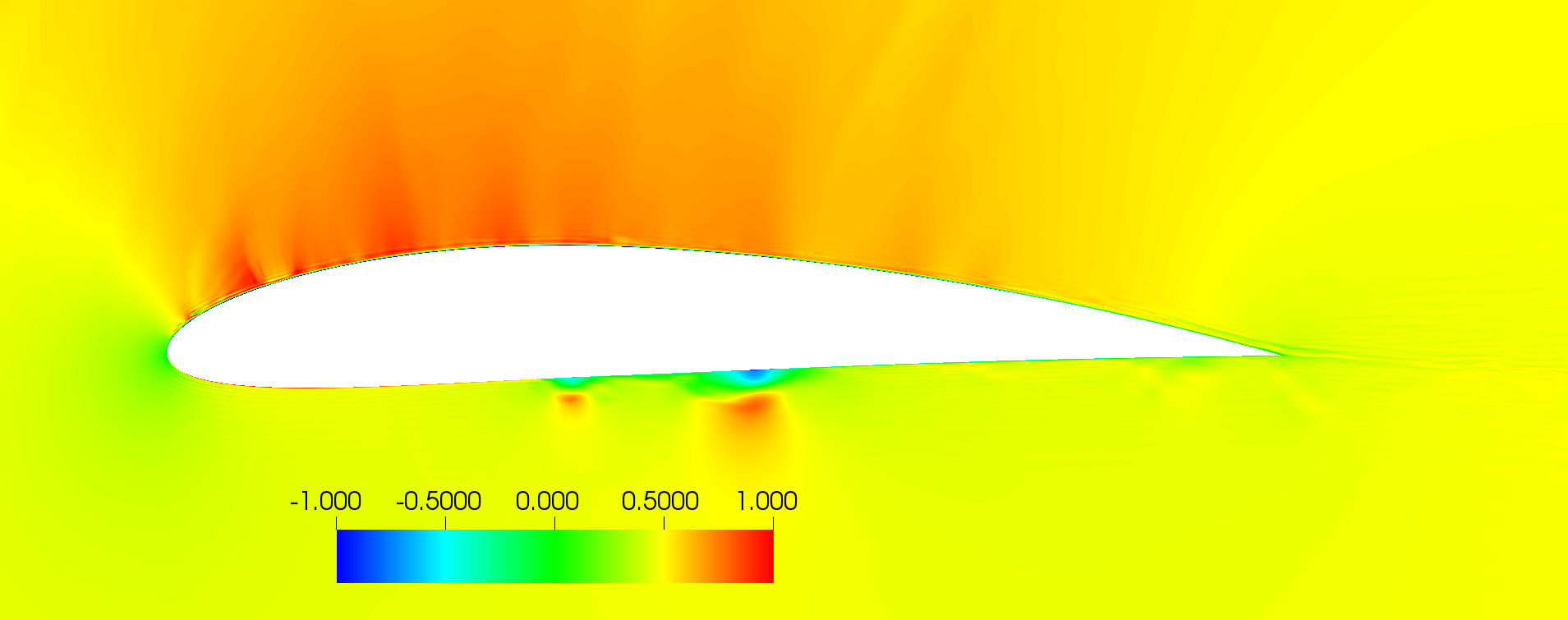}
	 \caption{Horizontal velocity contours over NACA4412 with equilibrium stress wall model boundary condition.
     Note the unstable flow over both surfaces.}
	 \label{f:blow_n4412}
	 \end{center}
	\end{figure}
	 \begin{figure}[hbt!]
	 \begin{center}
	   \includegraphics[trim = {0cm 0cm 0cm 0cm}, 
	   					clip, height = 6cm, width = 14cm]{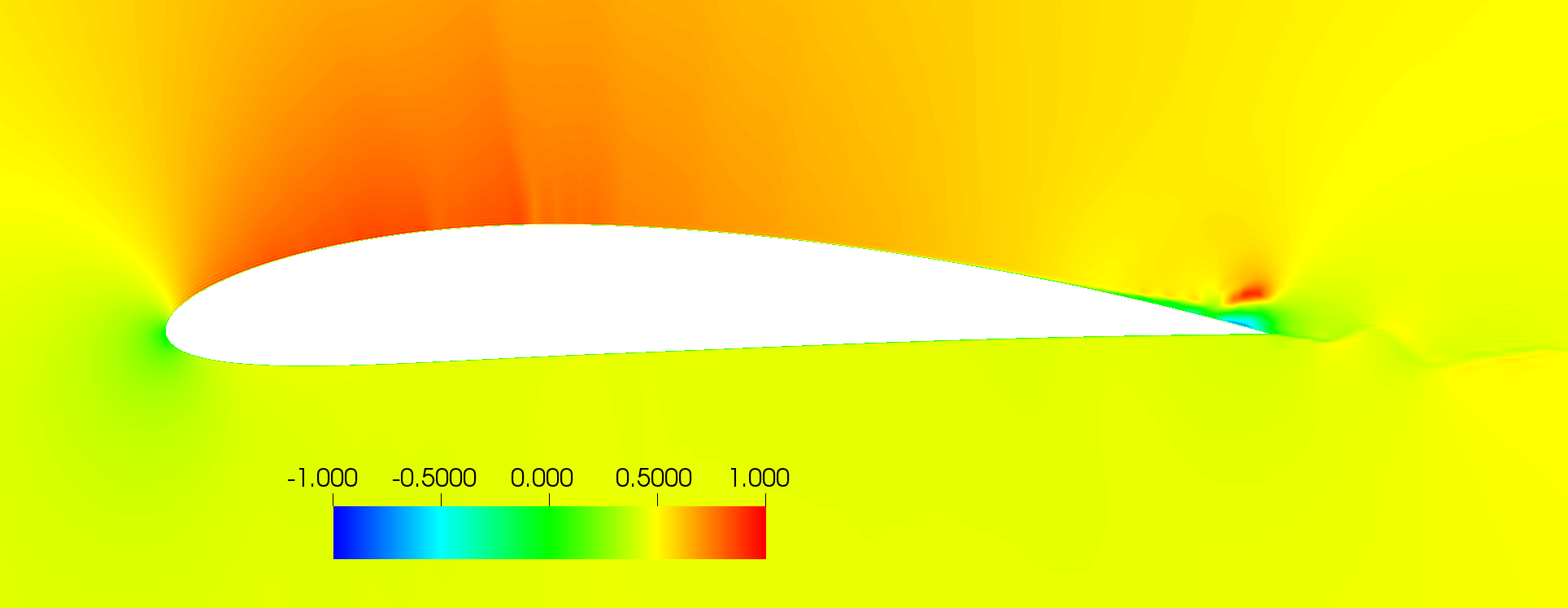}
	 \caption{Horizontal velocity contours over NACA4412 with dynamic slip boundary condition.}
	 \label{f:stable_n4412}
	 \end{center}
	\end{figure}
	 \begin{figure}[hbt!]
	 \begin{center}
	   \includegraphics[trim = {0cm 0cm 0cm 0cm}, 
	   					clip, height = 6cm, width = 11cm]{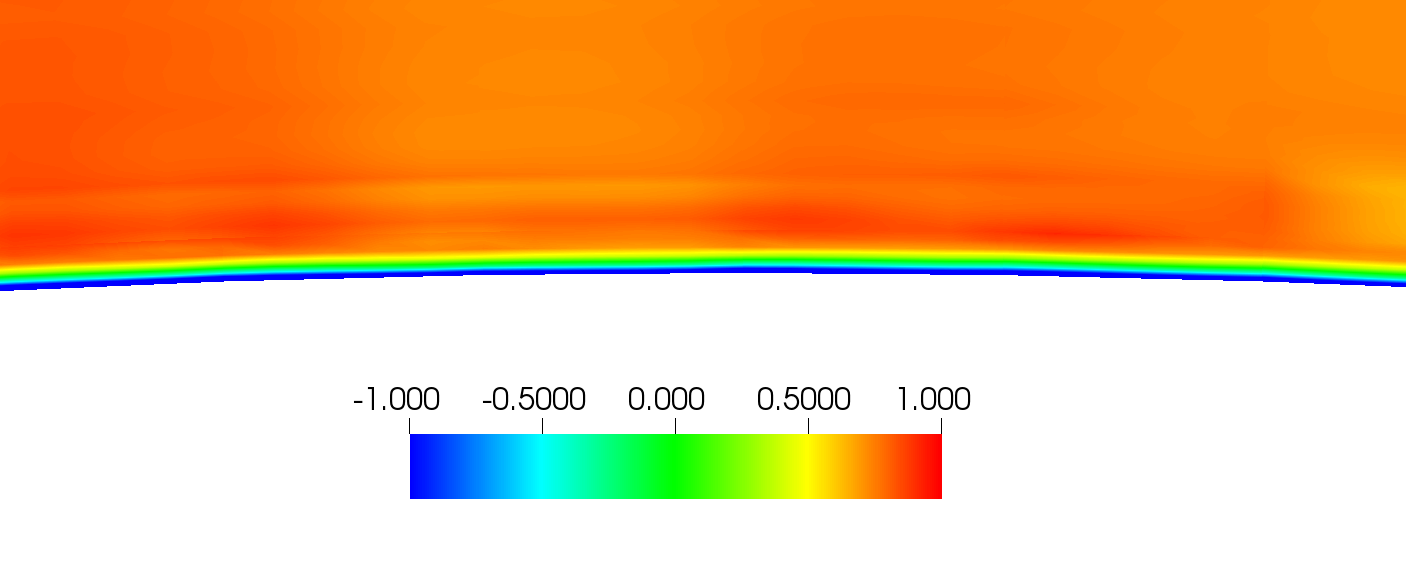}
	 \caption{Close-up of horizontal velocity contours over NACA4412 with equilibrium stress wall model boundary condition.
     Note the high negative velocities near the wall.}
	 \label{f:blow_n4412_close}
	 \end{center}
	\end{figure}
	 \begin{figure}[hbt!]
	 \begin{center}
	   \includegraphics[trim = {0cm 0cm 0cm 0cm}, 
	   					clip, height = 6cm, width = 11cm]{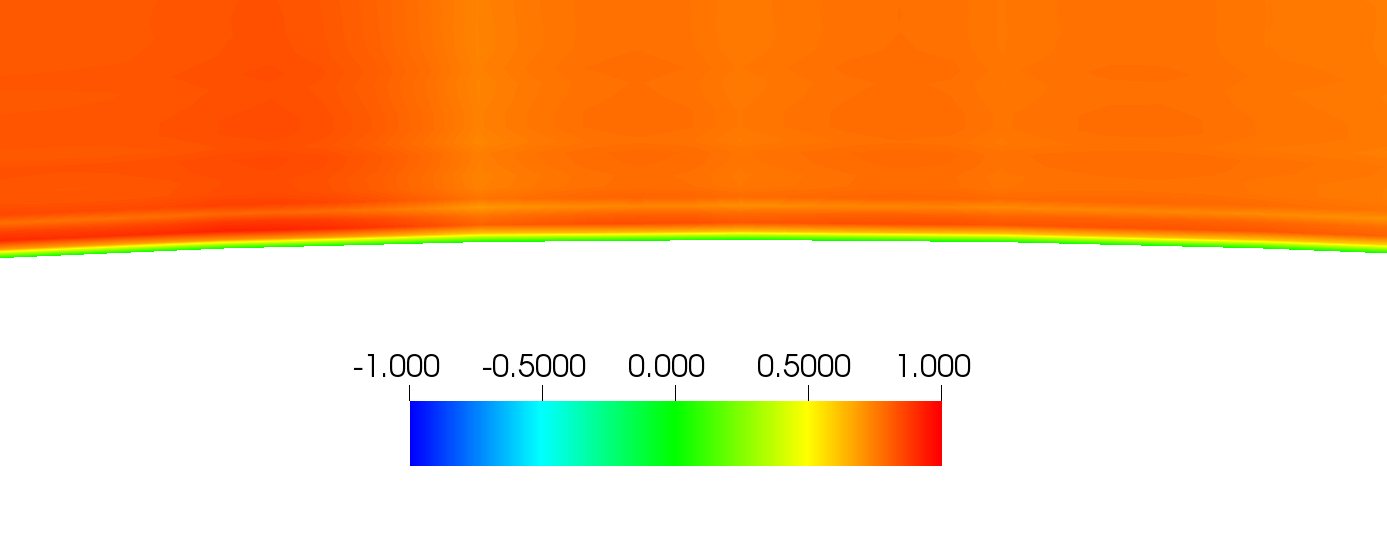}
	 \caption{Close-up horizontal velocity contours over NACA4412 with dynamic slip wall model boundary condition.}
	 \label{f:stable_n4412_close}
	 \end{center}
	\end{figure}

The NACA4412 airfoil is a popular test case for 
various experimental and numerical studies \cite{Coles1979, Hosseini2016, Frre2018}. 
Flow over an airfoil is characterized by changing velocity 
and pressure gradients unlike in the channel flow.
We use this test case to demonstrate that 
there are stability problems with the wall model
as our analysis indicates.
We have previously shown the effectiveness of using the equilibrium wall model
for this airfoil \cite{Singh2020}.
The wall model was used for flow at $Re \approx 1.64 \times 10^6, M=0.2, \alpha = 12^\circ$.
In this paper, we want to show here that with increasing Reynolds' and Mach number,
stability issues can make the equilibrium wall model unsuitable. 
To show this, we take a 2D unstructured grid and simulate flow at 
$Re \approx 4.0 \times 10^6, M=0.5, \alpha = 5^\circ$, 
where $\alpha$ is the angle of attack.
The mach number and angle of attack are only moderately high and 
so should not be challenging for a stable solver.

We ran the simulation to test robustness.
With the equilibrium stress wall model, the simulation only ran for 4 flow-through times and then blew-up.
Whereas for the dynamic slip wall model, the simulation did not blow up even after 30 flow through time periods.
Figures \ref{f:blow_n4412} and \ref{f:stable_n4412} show horizontal velocity over the airfoil for the two wall models. 
For the equilibrium stress wall model, we see irregular flow features all over the surface in contrast to the dynamic slip wall model.
A more detailed illustration of these issues can be seen in figures \ref{f:blow_n4412_close}
 and \ref{f:stable_n4412_close}. 
For the equilibrium stress wall model, we observe very high negative velocity near the wall. 
On the other hand, for the dynamic slip wall model, the near wall velocities are close to 0 .

The high negative velocity in the equilibrium stress wall model occurs since the shear stress is not guaranteed to oppose the velocity at the wall, 
but rather a velocity that is away from the wall. 
This leads to kinetic energy increase and eventually to blow up.

\subsection{Hybrid wall model} 

We have also investigated a hybrid wall model which 
choses one of the two wall models based on a particular condition.
This condition guarantees kinetic energy stability.
\begin{equation}\label{eq:hyb_bc}
        \tau^*_{xy}= 
        \begin{cases}
              {u}_{wm, x}\frac{\tau_w}{|\underline{u}_{wm}|}
            ,& \text{if } u \cdot u_{wm, x} > 0 
            \\
            \tau_{xy} + \sigma (u - l_p \frac{\partial u}{\partial y})
                ,              & \text{otherwise}
        \end{cases}
\end{equation}
where $u, {u}_{wm, x}$ are the components of the velocity in the 
$x$ direction at $y=0, \ y=y_{wm}$ respectively.
Note that equation (\ref{eq:hyb_bc}) selects (\ref{eq:slip_bc}) 
or (\ref{eq:reic1}) depending on the condition for the velocities. 
Shear stress boundary values for $\tau^*_{yz}$ can be similarly chosen.

%
%
\begin{figure}[hbt!]
    \begin{center}
        \includegraphics[trim = {0cm 0cm 0cm 0cm}, 
          clip, height = 10cm, width = 14cm]{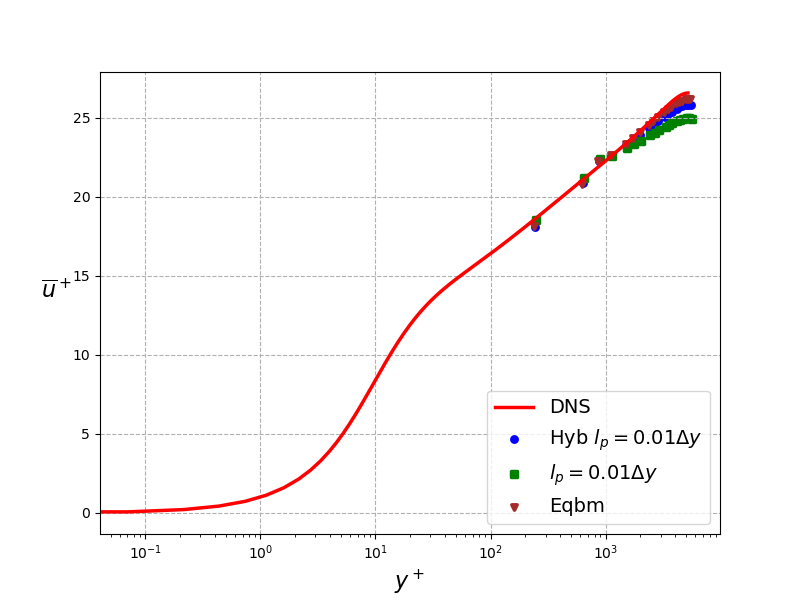}
        \caption{Time averaged mean profiles of streamwise velocity at $Re_\tau$ = 5200 
        for wall modelled ILES (WMILES) using all three wall models compared with DNS from \cite{moser_2015}.}
        \label{f:tcf_5200_wm_3}
    \end{center}
\end{figure}
\begin{figure}[hbt!]
    \begin{center}
        \includegraphics[trim = {0cm 0cm 0cm 0cm}, 
          clip, height = 10cm, width = 14cm]{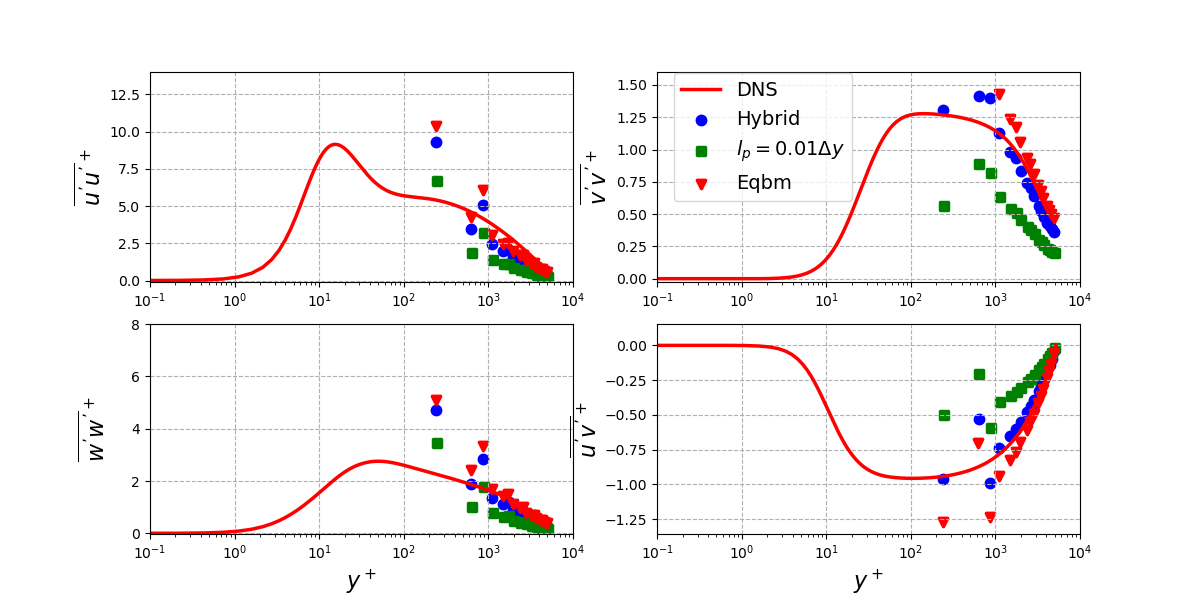}
        \caption{Time averaged mean profiles of second moments at $Re_\tau$ = 5200 
        for wall modelled ILES (WMILES) using all three wall models compared with DNS from \cite{moser_2015}.}
        \label{f:tcf_5200_var_hyb}
    \end{center}
\end{figure}
Figure \ref{f:tcf_5200_wm_3} compares the mean velocity profiles 
for the three wall models against DNS data.
We select the parameter $l_p = 0.01 \Delta y$ since that was the closest
to the DNS profile for the dynamic slip wall model.
The hybrid wall model curve is much closer to the DNS curve than
the dynamic slip wall model. 
However, the equilibrium wall model is still more accurate.
Figure \ref{f:tcf_5200_var_hyb} compares the second moments.
Again, we observe similar features as the mean velocity.
This suggests that using a hybrid wall model is a possible 
solution to a stable and accurate wall model.
For the sheared flow test case, this wall model behaves very similar
to the dynamic slip wall model.


\section{Conclusion}\label{sec:conc}

In the present study we investigate stability of wall modelled DG for 
the compressible Navier Stokes equations.
First, the formulations and the algorithm for implementing the wall model 
have been introduced. 
Next, kinetic energy stability estimates in the continuous case were developed, 
which show that the wall shear stress must oppose the velocity at the wall.
Further, we show that this estimate can be extended to the 
discrete DGSEM formulation because of its SBP property. 
This framework is used to investigate two wall models.
The popular equilibrium stress wall model is not kinetic energy stable, 
whereas the dynamic slip wall model is stable. 

Next, two test cases are studied numerically
to contrast the accuracy and stability properties of the wall models.
First, we use the turbulent channel flow to show that at high Reynolds' number, 
the first moment is accurately captured by the equilibrium stress wall model 
away from the wall.
However, the dynamic slip wall model is not very accurate.
Next, the NACA4412 test case is run at high Reynolds; 
number but at moderate mach number and angle of attack.
With the equilibrium stress wall model, 
the flow becomes very irregular and blows-up very quickly.
Near the wall, there are very large negative velocities. 
This occurs because the wall model does not satisfy the kinetic energy stability estimate.
However, with the dynamic slip wall model, the flow is smooth and stable.
Finally, a hybrid wall model is shown that choses 
the equilibrium wall model when it is stable and 
the dynamic slip wall model otherwise.
This is shown to behave much better in terms of accuracy to the dynamic slip wall mdoel.
This shows a possible way forward towards accurate and stable wall modeling.
To enable wider use of the kinetic energy stable dynamic slip wall model, 
better parameterizations are required for more accuracy.

\section*{Acknowledgments}
Vikram Singh and Steven Frankel would like to acknowledge 
that this work was partially supported by Israel Science
Foundation ISF-NSFC joint research program (ISF Grant No. 2232/15).
Jan Nordström was supported by Vetenskapsrådet (award number 2018-05084 VR), Sweden.

\appendix

\section{Discrete kinetic energy estimate} \label{ap:ke}

Here, we derive a discrete estimate for the kinetic energy for the viscous terms.
Using (\ref{eq:tau2}) in (\ref{eq:dis_ke}) yields
\begin{equation} \label{eq:ke_tau}
\begin{aligned}
\frac{d }{d t} \underline{1}^T  \doubleunderline{M} \ \underline{k}
&=
\underline{u}^T \doubleunderline{M}
\doubleunderline{D_1} \ \underline{ \tau_{xx} }  
+
\underline{u}^T  \doubleunderline{R_1}^T  \doubleunderline{B_1}
(
\underline{ \tau_{xx}^{*} }- \doubleunderline{R_1} \ \underline{\tau_{xx}}
)
\\
&+
\underline{u}^T \doubleunderline{M}
\doubleunderline{D_2} \ \underline{ \tau_{xy} }  
+
\underline{u}^T  \doubleunderline{R_2}^T  \doubleunderline{B_2}
(
\underline{ \tau_{xy}^{*} }- \doubleunderline{R_2} \ \underline{\tau_{xy}}
)
\\
&+
\underline{v}^T \doubleunderline{M}
\doubleunderline{D_1} \ \underline{ \tau_{xy} }  
+
\underline{v}^T  \doubleunderline{R_1}^T  \doubleunderline{B_1}
(
\underline{ \tau_{xy}^{*} }- \doubleunderline{R_1} \ \underline{\tau_{xy}}
)
\\
&+
\underline{v}^T \doubleunderline{M}
\doubleunderline{D_2} \ \underline{ \tau_{yy} }  
+
\underline{v}^T  \doubleunderline{R_2}^T  \doubleunderline{B_2}
(
\underline{ \tau_{yy}^{*} }- \doubleunderline{R_2} \ \underline{\tau_{yy}}
)
\\
&=
\underline{ \tau_{xx} } ^T \doubleunderline{D_1}^T
\doubleunderline{M}\ \underline{u}
+
\underline{u}^T  \doubleunderline{R_1}^T  \doubleunderline{B_1}
(
\underline{ \tau_{xx}^{*} }- \doubleunderline{R_1} \ \underline{\tau_{xx}}
)
\\
&+
\underline{ \tau_{xy} } ^T \doubleunderline{D_2}^T
\doubleunderline{M}\ \underline{u}
+
\underline{u}^T  \doubleunderline{R_2}^T  \doubleunderline{B_2}
(
\underline{ \tau_{xy}^{*} }- \doubleunderline{R_2} \ \underline{\tau_{xy}}
)
\\
&+
\underline{ \tau_{xy} } ^T \doubleunderline{D_1}^T
\doubleunderline{M}\ \underline{v}
+
\underline{v}^T  \doubleunderline{R_1}^T  \doubleunderline{B_1}
(
\underline{ \tau_{xy}^{*} }- \doubleunderline{R_1} \ \underline{\tau_{xy}}
)
\\
&+
\underline{ \tau_{yy} } ^T \doubleunderline{D_2}^T
\doubleunderline{M}\ \underline{v}
+
\underline{v}^T  \doubleunderline{R_2}^T  \doubleunderline{B_2}
(
\underline{ \tau_{yy}^{*} }- \doubleunderline{R_2} \ \underline{\tau_{yy}}
).
\end{aligned}
\end{equation}
since terms like
$
\underline{u}^T \doubleunderline{M}
\doubleunderline{D_1} \ \underline{ \tau_{xx} } 
$
are scalars implying
$
\underline{u}^T \doubleunderline{M}
\doubleunderline{D_1} \ \underline{ \tau_{xx} } 
=
\underline{ \tau_{xx} } ^T \doubleunderline{D_1}^T
\doubleunderline{M}\ \underline{u}. 
$
If we multiply the first equation in (\ref{eq:tau1}) with
$\underline{\tau_{xx}}^T \doubleunderline{M}$ and the second with 
$\underline{\tau_{yy}}^T \doubleunderline{M}$ from the left, we get
\begin{equation} \label{eq:tau3}
\begin{aligned}
\underline{\tau_{xx}}^T \doubleunderline{M}
(
\underline{ \tau_{xx} }
+
\frac{1}{2}
\underline{ \tau_{yy} }
)
&=
\mu 
\underline{\tau_{xx}}^T \doubleunderline{M}
\doubleunderline{D_1} \ \underline{ u }  
+
\mu 
\underline{\tau_{xx}}^T 
\doubleunderline{R_1}^T  \doubleunderline{B_1}
(\underline{u}^* - \doubleunderline{R_1} \  \underline{u})
\\
\underline{\tau_{yy}}^T \doubleunderline{M}
(
\underline{ \tau_{yy} }
+
\frac{1}{2}
\underline{ \tau_{xx} }
)
&=
\mu 
\underline{\tau_{xx}}^T \doubleunderline{M}
\doubleunderline{D_2} \ \underline{ v }  
+
\mu 
\underline{\tau_{xx}}^T 
 \doubleunderline{R_2}^T  \doubleunderline{B_2}
(\underline{v}^* - \doubleunderline{R_2} \  \underline{v}).
\end{aligned}
\end{equation}
Using the SBP property (\ref{eq:sbp_d}) in (\ref{eq:tau3}) yields
\begin{equation} \label{eq:tau_xx}
\begin{aligned}
\frac{1}{\mu}
\underline{\tau_{xx}}^T \doubleunderline{M}
(
\underline{ \tau_{xx} }
+
\frac{1}{2}
\underline{ \tau_{yy} }
)
&=
-
\underline{\tau_{xx}}^T \doubleunderline{D_1}^T
\doubleunderline{M} \ \underline{ u }  
+
\underline{\tau_{xx}}^T 
\doubleunderline{R_1}^T  \doubleunderline{B_1} \ \underline{u}^* 
\\
\frac{1}{\mu}
\underline{\tau_{yy}}^T \doubleunderline{M}
(
\underline{ \tau_{yy} }
+
\frac{1}{2}
\underline{ \tau_{xx} }
)
&=
-
\underline{\tau_{xx}}^T \doubleunderline{D_2}^T
\doubleunderline{M} \ \underline{ v }  
+
\underline{\tau_{xx}}^T 
 \doubleunderline{R_2}^T  \doubleunderline{B_2} \ \underline{v}^*.
\end{aligned}
\end{equation}
Similarly, multiplying the expression for $\tau_{xy}$ on the left by 
$\underline{ \tau_{xy} }^T\doubleunderline{M}$ in (\ref{eq:tau}) yields
\begin{equation} \label{eq:tau4}
\begin{aligned}
\frac{1}{\mu}
\underline{ \tau_{xy} }^T
\doubleunderline{M}
\underline{ \tau_{xy} }
&=
\underline{ \tau_{xy} }^T
\doubleunderline{M}
\doubleunderline{D_2} \ \underline{ u } + 
\underline{ \tau_{xy} }^T
\doubleunderline{M}
\doubleunderline{D_1} \ \underline{ v }
+
\underline{ \tau_{xy} }^T
\doubleunderline{R_1}^T  \doubleunderline{B_1}
(\underline{v}^* - \doubleunderline{R_1} \  \underline{v})
\\
&+
\underline{ \tau_{xy} }^T
\doubleunderline{R_2}^T  \doubleunderline{B_2}
(\underline{u}^* - \doubleunderline{R_2} \  \underline{u})
\end{aligned}
\end{equation}
Again using the SBP property (\ref{eq:sbp_d}) in (\ref{eq:tau4}) yields
\begin{equation} \label{eq:tau_xy}
\begin{aligned}
\frac{1}{\mu}
\underline{ \tau_{xy} }^T
\doubleunderline{M}
\underline{ \tau_{xy} }
&=
-\underline{ \tau_{xy} }^T \doubleunderline{D_2}^T 
\doubleunderline{M} \ \underline{ u } 
- 
\underline{ \tau_{xy} }^T \doubleunderline{D_1} ^T
\doubleunderline{M} \ \underline{ v }
+
\underline{ \tau_{xy} }^T
\doubleunderline{R_1}^T  \doubleunderline{B_1} \ \underline{v}^*
+
\underline{ \tau_{xy} }^T
\doubleunderline{R_2}^T  \doubleunderline{B_2} \underline{u}^* 
\end{aligned}
\end{equation}
Using (\ref{eq:tau_xx}) and (\ref{eq:tau_xy}) in the expression for kinetic energy (\ref{eq:ke_tau}) yields
\begin{equation} \label{eq:ke_tau1}
\begin{aligned}
\frac{d }{d t} \underline{1}^T  \doubleunderline{M} \ \underline{k}
&=
-
\frac{1}{\mu}
\underline{ \tau_{xy} }^T
\doubleunderline{M}
\underline{ \tau_{xy} }
+
\underline{ \tau_{xy} }^T
\doubleunderline{R_1}^T  \doubleunderline{B_1} \ \underline{v}^*
+
\underline{ \tau_{xy} }^T
\doubleunderline{R_2}^T  \doubleunderline{B_2} \underline{u}^* 
\\
&-
\frac{1}{\mu}
\underline{\tau_{xx}}^T \doubleunderline{M}
(
\underline{ \tau_{xx} }
+
\frac{1}{2}
\underline{ \tau_{yy} }
)
-
\frac{1}{\mu}
\underline{\tau_{yy}}^T \doubleunderline{M}
(
\underline{ \tau_{yy} }
+
\frac{1}{2}
\underline{ \tau_{xx} }
)
\\
&+
\underline{\tau_{xx}}^T 
\doubleunderline{R_1}^T  \doubleunderline{B} \ \underline{u}^*
+
\underline{\tau_{yy}}^T 
\doubleunderline{R_2}^T  \doubleunderline{B} \ \underline{v}^*
\\
&+
\underline{u}^T  \doubleunderline{R_1}^T  \doubleunderline{B_1}
(
\underline{ \tau_{xx}^{*} }- \doubleunderline{R_1} \ \underline{\tau_{xx}}
)
+
\underline{u}^T  \doubleunderline{R_2}^T  \doubleunderline{B_2}
(
\underline{ \tau_{xy}^{*} }- \doubleunderline{R_2} \ \underline{\tau_{xy}}
)
\\
&+
\underline{v}^T  \doubleunderline{R_1}^T  \doubleunderline{B_1}
(
\underline{ \tau_{xy}^{*} }- \doubleunderline{R_1} \ \underline{\tau_{xy}}
)
+
\underline{v}^T  \doubleunderline{R_2}^T  \doubleunderline{B_2}
(
\underline{ \tau_{yy}^{*} }- \doubleunderline{R_2} \ \underline{\tau_{yy}}
).
\end{aligned}
\end{equation}
Using the property that $\doubleunderline{M}$ is a positive definite matrix and the identity (\ref{eq:iden}) yields
\begin{equation} \label{eq:tau_pos}
\begin{aligned}
\frac{1}{\mu}
\underline{\tau_{xx}}^T \doubleunderline{M}
(
\underline{ \tau_{xx} }
+
\frac{1}{2}
\underline{ \tau_{yy} }
)
+
\frac{1}{\mu}
\underline{\tau_{yy}}^T \doubleunderline{M}
(
\underline{ \tau_{yy} }
+
\frac{1}{2}
\underline{ \tau_{xx} }
)
+
\frac{1}{\mu}
\underline{ \tau_{xy} }^T
\doubleunderline{M}
\underline{ \tau_{xy} }
=
\\
\frac{1}{\mu}
(
\underline{ \tau_{xy} }^T
\doubleunderline{M}
\underline{ \tau_{xy} }
+
\underline{\tau_{xx}}^T \doubleunderline{M}
\underline{ \tau_{xx} }
+
\underline{\tau_{xx}}^T \doubleunderline{M}
\underline{ \tau_{yy} }
+
\underline{\tau_{yy}}^T \doubleunderline{M}
\underline{ \tau_{yy} }
)
\geq 0
\end{aligned}
\end{equation}
Using (\ref{eq:tau_pos}) in (\ref{eq:ke_tau1}) yields
\begin{equation} \label{eq:ke_tau2}
\begin{aligned}
\frac{d }{d t} \underline{1}^T  \doubleunderline{M} \ \underline{k}
&\leq
\underline{ \tau_{xy} }^T
\doubleunderline{R_1}^T  \doubleunderline{B_1} \ \underline{v}^*
+
\underline{ \tau_{xy} }^T
\doubleunderline{R_2}^T  \doubleunderline{B_2} \ \underline{u}^* 
+
\underline{\tau_{xx}}^T 
\doubleunderline{R_1}^T  \doubleunderline{B_1} \ \underline{u}^*
+
\underline{\tau_{yy}}^T 
\doubleunderline{R_2}^T  \doubleunderline{B_2} \ \underline{v}^*
\\
&+
\underline{u}^T  \doubleunderline{R_1}^T  \doubleunderline{B_1}
(
\underline{ \tau_{xx}^{*} }- \doubleunderline{R_1} \ \underline{\tau_{xx}}
)
+
\underline{u}^T  \doubleunderline{R_2}^T  \doubleunderline{B_2}
(
\underline{ \tau_{xy}^{*} }- \doubleunderline{R_2} \ \underline{\tau_{xy}}
)
\\
&+
\underline{v}^T  \doubleunderline{R_1}^T  \doubleunderline{B_1}
(
\underline{ \tau_{xy}^{*} }- \doubleunderline{R_1} \ \underline{\tau_{xy}}
)
+
\underline{v}^T  \doubleunderline{R_2}^T  \doubleunderline{B_2}
(
\underline{ \tau_{yy}^{*} }- \doubleunderline{R_2} \ \underline{\tau_{yy}}
).
\end{aligned}
\end{equation}
As before, if we look at only the $y$ boundary (\ref{eq:ke_tau2}) becomes
\begin{equation} \label{eq:ke_tau_y}
\begin{aligned}
\frac{d }{d t} \underline{1}^T  \doubleunderline{M} \ \underline{k}
&\leq
\underline{ \tau_{xy} }^T
\doubleunderline{R_2}^T  \doubleunderline{B_2} \ \underline{u}^* 
+
\underline{\tau_{yy}}^T 
\doubleunderline{R_2}^T  \doubleunderline{B_2} \ \underline{v}^*
+
\underline{u}^T  \doubleunderline{R_2}^T  \doubleunderline{B_2}
(
\underline{ \tau_{xy}^{*} }- \doubleunderline{R_2} \ \underline{\tau_{xy}}
)
\\
&+
\underline{v}^T  \doubleunderline{R_2}^T  \doubleunderline{B_2}
(
\underline{ \tau_{yy}^{*} }- \doubleunderline{R_2} \ \underline{\tau_{yy}}
).
\end{aligned}
\end{equation}
Substituting the boundary condition for slip velocity  
$\underline{v}^* = \underline{0}, 
\underline{u}^* = \doubleunderline{R_2}\ \underline{u}$ in
(\ref{eq:ke_tau_y}) yields
\begin{equation} \label{eq:ke_tau_y}
\begin{aligned}
\frac{d }{d t} \underline{1}^T  \doubleunderline{M} \ \underline{k}
&\leq
\underline{ \tau_{xy} }^T
\doubleunderline{R_2}^T  \doubleunderline{B_2} \ \doubleunderline{R_2}\ \underline{u}
+
\underline{u}^T  \doubleunderline{R_2}^T  \doubleunderline{B_2}
(
\underline{ \tau_{xy}^{*} }- \doubleunderline{R_2} \ \underline{\tau_{xy}}
)
\\
&=
\underline{u}^T  \doubleunderline{R_2}^T  \doubleunderline{B_2} \
\underline{ \tau_{xy}^{*} }.
\end{aligned}
\end{equation}
%
%
%

\bibliographystyle{siam}
\bibliography{refs}

\end{document}